\documentclass[twocolumn,showpacs,amsmath,amssymb,prd,floatfix]{revtex4}

\usepackage{graphicx}
\usepackage{dcolumn}
\usepackage{bm}
\usepackage[varg]{txfonts}

\newcommand{\threej}[6]{
\left(
    \begin{array}{ccc}
       \! #1 \! & \! #2 \! & \! #3 \!\! \\
       \! #4 \! & \! #5 \! & \! #6 \!\!
    \end{array}
\right)
}
\newcommand{\sthreej}[6]{
\left({}^{#1}_{#4}{}^{#2}_{#5}{}^{#3}_{#6}\right)
}
\newcommand{\sixj}[6]{
\left\{
    \begin{array}{ccc}
       \! #1 \!\! & \!\! #2 \!\! & \!\! #3 \!\!\! \\
       \! #4 \!\! & \!\! #5 \!\! & \!\! #6 \!\!\!
    \end{array}
\right\}
}
\newcommand{\ssixj}[6]{
\left\{{}^{#1}_{#4}{}^{#2}_{#5}{}^{#3}_{#6}\right\}
}

\newcommand{\sumst}[3]{
  \sum_{\mbox{\scriptsize
      $\begin{array}{c} #1 \\ #2 \\ #3 \end{array}$}}
}

\begin{document}

\title{

  Analytic Minkowski Functionals of the Cosmic Microwave Background:\\
  Second-order Non-Gaussianity with Bispectrum and Trispectrum

}

\author{Takahiko Matsubara}
\email{taka@a.phys.nagoya-u.ac.jp}
\affiliation{%
Department of Physics, Nagoya University,
Chikusa, Nagoya, 464-8602, Japan
}%

\date{\today}

\begin{abstract}
    Analytic formulas of Minkowski functionals in two-dimensional
    random fields are derived, including effects of second-order
    non-Gaussianity in the presence of both the bispectrum and
    trispectrum. The set of formulas provides a promising method to
    constrain the primordial non-Gaussianity of the universe by
    temperature fluctuations in the cosmic microwave background
    radiation. In a case of local-type non-Gaussianity, the Minkowski
    functionals are analytically given by powers of quadratic and
    cubic parameters, $f_{\rm NL}$ and $g_{\rm NL}$. Our formulas are
    not restricted to this particular model, and applicable to a wide
    class of non-Gaussian models. The analytic formulas are compared
    to numerical evaluations from non-Gaussian realizations of
    temperature maps, showing very good agreements.
\end{abstract}

\maketitle


\section{\label{sec:intro}
Introduction
}

Detecting the primordial non-Gaussianity will play a key role in
discriminating the models of the early universe. All models of
inflation predict the primordial non-Gaussianity to some extent in
principle. While a simple inflationary model with a single
slow-rolling scalar field produces too small non-Gaussianity to be
observed \cite{muk81,sta82,haw82,gut82,all87,gan94}, many other models
such as the curvaton scenario \cite{lyt02,lin97,lyt03}, models with
non-standard kinetic term \cite{arm99,gar99}, cyclic or ekpyrotic
universes without inflation \cite{kho01}, etc.~predict large
non-Gaussianity which could be detected in any future \cite{bar04}.

The statistical properties of random Gaussian fields are completely
characterized by the two-point correlation function in configuration
space, or the power spectrum in Fourier space. Thus the information of
non-Gaussianity is contained in higher-order polyspectra, such as the
bispectrum, trispectrum, and so forth. Recently, much work is devoted
to analyzing the temperature fluctuations in the cosmic microwave
background (CMB) to constrain the primordial non-Gaussianity
\cite{kom03,spe07,cre07,kom09,smi09,cur09}. So far most of the
constraints are placed to a phenomenological parameter $f_{\rm NL}$,
which is largely responsible for the primordial bispectrum. However,
the bispectrum is not sufficient to fully discriminate the wide
variety of models of the early universe. Some models, such as the
curvaton scenario \cite{sas06,hua09,byr09}, ekpyrotic universes
\cite{leh09}, etc.~can generate a large trispectrum with a relatively
small bispectrum. The trispectrum is often characterized by
phenomenological parameters $g_{\rm NL}$ (and $\tau_{\rm NL}$ in some
models). When $g_{\rm NL}$ is large enough such that $g_{\rm NL} \gg
f_{\rm NL}^2$, the bispectrum is not significant and the primordial
non-Gaussianity can be detected only through the effect of
trispectrum.

Straightforward measurements of the polyspectra become progressively
complicated for higher-order statistics beyond the bispectrum, since
the higher-order polyspectra have many arguments with complicated
dependence on shapes and scales. Even though direct calculations of
the higher-order polyspectra may be primary methods to constrain
higher-order non-Gaussianity, it is desirable to have as many
alternative methods as possible. Among others, the set of Minkowski
functionals (MFs) \cite{mec94,sch97,sch98} has been proved to be a
simple and powerful tool in analyzing non-Gaussianity in CMB
\cite{nov00,wu01,pol02,kom03,spe07,det07,cur07,cur08,kom09}.

Since the hierarchy of polyspectra contains all the information of the
statistical property of random field, the expectation values of MFs
should be expressible by the polyspectra. Such relations were
analytically derived for weakly non-Gaussian fields in the
lowest-order approximation \cite{mat03}. It was shown that deviations
of the non-Gaussian MFs from the Gaussian MFs are proportional to a
linear combination of the bispectrum. The analytic formulas are
incorporated with the CMB bispectrum \cite{hik06} and are applied to
the analysis of CMB, successfully constraining non-Gaussianity
parameter $f_{\rm NL}$ without relying on massive numerical
simulations to generate non-Gaussian CMB maps
\cite{hik08,nat09,hik09}.

So far the analytic formulas of non-Gaussian MFs are known only in the
lowest-order approximation, which we call first-order non-Gaussianity.
The contributions from the trispectrum and other polyspectra to the
geometrical descriptors, such as the MFs, appear in higher-order
non-Gaussianities \cite{mat03,chi09,pog09}. To extract any information
from MFs beyond the bispectrum, such as $g_{\rm NL}$ and $\tau_{\rm
  NL}$, it is particularly useful to have analytic formulas with
higher-order approximations. The purpose of this paper is to derive
the analytic formulas with second-order approximation of
non-Gaussianity, including the effects up to the trispectrum. These
primary formulas are applicable to a wide class of non-Gaussian fields
in 2-dimensional space. We also focus on how the formulas are applied
to analyses of CMB temperature fluctuations. Assuming a local model of
non-Gaussianity, we derive the explicit dependence of the MFs on
parameters $f_{\rm NL}$ and $g_{\rm NL}$ (and $\tau_{\rm NL}$).

This paper is organized as follows: In \S 2, the analytic formulas of
MFs with second-order non-Gaussianity are derived. Non-Gaussianity
parameters to describe the analytic MFs are explicitly given by the
bispectrum and trispectrum in \S 3. In \S 4, the polyspectra in the
local model of non-Gaussianity are summarized for applications to the
analytic MFs. Numerical tests of the analytic formulas are given in \S
5, and our results are summarized in \S 6.

\section{\label{sec:formula}
Analytic Minkowski Functionals with Second-order Non-Gaussianity
}

The MFs are morphological descriptors to study statistical properties
of random fields. For a smooth scalar field $f(\bm{x})$ with
zero mean, $\langle f \rangle = 0$, the excursion set over a given
threshold $\nu$, $Q = \{\bm{x} | f(\bm{x}) > \nu \sigma\}$ are
defined, where $\sigma \equiv \langle f^2 \rangle^{1/2}$ is the rms of
the field. The MFs for a set $Q$ in a 2-dimensional space with smooth
boundary $\partial Q$ are given by \cite{sch98}
\begin{equation}
  V_0(\nu) = \int_Q da, \quad
  V_1(\nu) = \frac14 \int_{\partial Q} d\ell, \quad
  V_2(\nu) = \frac{1}{2\pi} \int_{\partial Q} \kappa d\ell,
\label{eq:01}
\end{equation}
where $da$ and $d\ell$ denote the surface element of $Q$ and the line
element along $\partial Q$, respectively, and $\kappa$ is the geodesic
curvature on the boundary $\partial Q$. Apart from numerical factors,
Minkowski functionals $V_0$, $V_1$, $V_2$ correspond to the area of
$Q$, the length of $\partial Q$, and the curvature integral of
$\partial Q$, respectively. According to the Gauss-Bonnet theorem,
$2\pi$ times the Euler characteristic $\chi$ of $Q$ is equal to the
integral of geodesic curvature of $\partial Q$ plus the integral of
Gaussian curvature of $Q$. Therefore, when the excursion set is
defined on the sphere of radius $R$, the Euler characteristic is given
by a linear combination of the Minkowski functionals via $\chi = V_2 +
V_0/2\pi R^2$ \cite{sch98}. In a flat space ($R \rightarrow \infty$),
$\chi = V_2$.

The expectation values of MFs in 2D are generically given by
analytic forms \cite{mat03}
\begin{equation}
  V_k(\nu) = 
   \frac{1}{(2\pi)^{(k+1)/2}}
   \frac{\omega_2}{\omega_{2-k}\omega_k}
   \left(\frac{\sigma_1}{\sqrt{2}\sigma}\right)^k
   e^{-\nu^2/2} v_k(\nu),
\label{eq:02}
\end{equation}
where $\sigma_1 \equiv \langle(\nabla u)^2 \rangle^{1/2}$ is the
variance of the gradient field, and $\omega_k \equiv
\pi^{k/2}/\Gamma(k/2 + 1)$ is the volume of the unit ball in $k$
dimensions, i.e., $\omega_0 = 1$, $\omega_1 = 2$, $\omega_2 = \pi$.
For a random Gaussian field, the MFs are analytically given by
Tomita's formula \cite{tom86}, which corresponds to
\begin{equation}
  v_k(\nu) = H_{k-1}(\nu),
\label{eq:02-1}
\end{equation}
where
\begin{equation}
    H_n(\nu) = e^{\nu^2/2} \left(-\frac{d}{d\nu}\right)^n e^{-\nu^2/2},
\label{eq:02-2}
\end{equation}
are the Hermite polynomials, and we define a function
\begin{equation}
    H_{-1}(\nu) \equiv
    e^{\nu^2/2}\int_\nu^\infty d\nu e^{-\nu^2/2}
    = \sqrt{\frac{\pi}{2}} {\rm erfc}\left(\frac{\nu}{\sqrt{2}}\right),
\label{eq:02-3}
\end{equation}
for $k=0$ in Eq.~({\ref{eq:02-1}).

  Throughout this paper, we assume hierarchical orderings of the
  higher-order correlators, $\langle f^n \rangle_{\rm c} \sim {\cal
    O}(\sigma^{2n-2})$, where $\langle \cdots \rangle_{\rm c}$ denotes
  the connected part, or the cumulant. In such a case, the reduced MFs
  $v_k(\nu)$ are expanded by the variance $\sigma$ as
\begin{equation}
  v_k = v_k^{(0)} +  v_k^{(1)} \sigma +  v_k^{(2)} \sigma^2
  + \cdots
\label{eq:03}
\end{equation}
where $v_k^{(0)} = H_{k-1}(\nu)$ is the Gaussian contribution to the
MFs. Primordial non-Gaussianities generated by most of the
inflationary models satisfy the hierarchical orderings which we
assume. In the previous work \cite{mat03}, the first-order corrections
to the MFs in a non-Gaussian field are analytically derived:
\begin{equation}
    v_k^{(1)}(\nu) =
    \frac{S}{6} H_{k+2}(\nu)
    - \frac{k S_{\rm I}}{4} H_k(\nu)
    - \frac{k(k-1) S_{\rm II}}{4} H_{k-2}(\nu),
\label{eq:04}
\end{equation}
where
\begin{equation}
  S = \frac{\langle f^3 \rangle}{\sigma^4}, \quad
  S_{\rm I} =
  \frac{\langle f^2 \nabla^2 f \rangle}
    {\sigma^2\sigma_1^2}, \quad
  S_{\rm II} =
  \frac{2\langle |\nabla f|^2 \nabla^2 f\rangle}
    {\sigma_1^4},
\label{eq:05}
\end{equation}
are the skewness parameter and its derivatives.

Fundamental techniques to evaluate the next-order term, $v_k^{(2)}$,
are mostly found in \cite{mat03}. We outline below the higher-order
extensions of the calculation in \cite{mat03}. First we define the
normalized cumulants of the field derivatives,
\begin{equation}
  M^{(n)}_{\mu_1\cdots\mu_n} \equiv
  \sigma^{2-2n} \langle f_{\mu_1} \cdots f_{\mu_n} \rangle_{\rm c}
\label{eq:05-1}
\end{equation}
(which corresponds to the quantity denoted by
$\hat{M}^{(n)}_{\mu_1\cdots\mu_n}$ in \cite{mat03}), where $f_\mu$
denotes all the field derivatives, $(f_\mu) = (f, f_{;1}, f_{;2},
f_{;11}, f_{;22}, f_{;12}, \ldots)$, and ``$;$'' indicates the
covariant derivative. In the following we adopt a notation, $f_0 = f$,
$f_1 = f_{;1}$, $f_2 = f_{;2}$, $f_{11} = f_{;11}$, and the
corresponding index runs over $\mu = 0,1,2,11$. The
Eq.~(\ref{eq:05-1}) is of zero-th order in $\sigma$ for hierarchical
orderings. For any statistic which is locally defined by some function
$F$ of field derivatives $f_\mu$, the expectation value of the statistic
is expanded as (Eq.~(22) of \cite{mat03})
\begin{align}
   \langle F \rangle &= 
   F^{\rm G} +
   \frac{1}{6}\sum M^{(3)}_{\mu_1 \mu_2 \mu_3}
   F^{\rm G}_{\mu_1 \mu_2 \mu_3} \sigma
\nonumber\\
&\quad +
   \left[
      \frac{1}{24} \sum M^{(4)}_{\mu_1 \mu_2 \mu_3 \mu_4}
      F^{\rm G}_{\mu_1\mu_2\mu_3\mu_4}
   \right.
\nonumber\\
&\qquad \left.\, +
      \frac{1}{72} \sum M^{(3)}_{\mu_1\mu_2\mu_3}
      M^{(3)}_{\mu_4\mu_5\mu_6}
      F^{\rm G}_{\mu_1 \mu_2 \mu_3 \mu_4 \mu_5 \mu_6}
  \right] \sigma^2
\nonumber\\
&\quad + {\cal O}(\sigma^3),
\label{eq:06}
\end{align}
where $F^{\rm G} \equiv \langle F \rangle_{\rm G}$, $F^{\rm
  G}_{\mu_1\mu_2\cdots} \equiv
\langle \partial_{\mu_1} \partial_{\mu_2}\cdots F \rangle_{\rm G}$,
$\partial_\mu = \partial/\partial f_\mu$, and $\langle \cdots
\rangle_{\rm G}$ denotes the expectation value for the Gaussian field
which has the same two-point functions as the non-Gaussian field we
consider (see \cite{mat03} for detail). The first two terms on RHS of
Eq.~(\ref{eq:06}) correspond to the Gaussian contribution and
first-order non-Gaussian corrections, respectively.

For the MFs, the function $F$ is given by \cite{mat03}
\begin{equation}
  F =
  \begin{cases}
    \Theta(u - \nu), & (k=0), \\
    \frac{\pi}{8} \delta(u - \nu) |u_1|, & (k=1), \\
    - \frac12 \delta(u - \nu) \delta(u_1) |u_2| u_{11}, & (k=2), 
  \end{cases}
\label{eq:07}
\end{equation}
where $\Theta$, $\delta$ are the step function and the delta function,
respectively, and $u_\mu = f_\mu/\sigma$. The
quantities $F^{\rm G}_{\mu_1\mu_2\cdots}$ are explicitly calculated to
give \cite{mat03}
\begin{align}
  F^{\rm G}_{\mu_1\mu_2\cdots} &=
  \frac{e^{-\nu^2/2}}{(2\pi)^{(k+1)/2}}
  q^{k-l_1-l_2-2m}
\nonumber\\
 & \quad \times
  \begin{cases}
    H_{n-1}(\nu) \delta_{l_10} \delta_{l_20} \delta_{m0},
    & (k=0), \\
    \frac{\pi}{4} H_n(\nu) h_{l_1-2}
    \delta_{l_20} \delta_{m0},
    & (k=1), \\
    h_{l_1}h_{l_2-2}
    \left[H_{n+1}(\nu) \delta_{m0} - H_n(\nu) \delta_{m1}\right]
    & (k=2), 
  \end{cases}
\label{eq:08}
\end{align}
where $q \equiv \sigma_1/\sqrt{2}\sigma$, and $n,l_1,l_2,m$ are
numbers of $0$, $1$, $2$, $11$, respectively, in a set of
indices $\mu_1,\mu_2,\ldots$. The factor $h_l$ is defined by
\begin{equation}
   h_l = 
   \begin{cases}
       0, & (l:\mbox{odd}),\\
       (-2)^{l/2}\pi^{-1/2}\Gamma[(l+1)/2], & (l:\mbox{even}).
   \end{cases}
\label{eq:08-1}
\end{equation}
For example, $h_{-2} = 1$, $h_{-1} = 0$, $h_{0} = 1$, $h_{1} = 0$,
$h_{2} = -1$, $h_{3} = 0$, $h_{4} = 3$, etc.

The factor $M^{(3)}_{\mu_1\mu_2\mu_3}$ is expressible by the skewness
and its derivatives of Eq.~(\ref{eq:05}) from rotational symmetry
\cite{mat03}. The non-zero components are
\begin{align} 
&
  M^{(3)}_{000} = S, \quad
  M^{(3)}_{00(11)} = q^2S_{\rm I},
\nonumber\\
&
  M^{(3)}_{011} = M^{(3)}_{022}
  = -\frac{q^2}{2} S_{\rm I}, \quad
  M^{(3)}_{22(11)} = q^4 S_{\rm II}
\label{eq:09}
\end{align} 
and their permutations. For other combinations of the indices,
$M^{(3)}_{\mu_1\mu_2\mu_3} = 0$. Substituting Eq.~(\ref{eq:08}) and
Eq.~(\ref{eq:09}) into the second term in the RHS of
Eq.~(\ref{eq:06}), the first-order corrections to the MFs of
Eq.~(\ref{eq:04}) follow.

It is straightforward to extend the above calculation to derive
second-order corrections. We only need to calculate 4-point cumulants
$M^{(4)}_{\mu_1\mu_2\mu_3\mu_4}$. With similar considerations of
rotational symmetry in the case of 3-point cumulants, we obtain
\begin{align} 
&
  M^{(4)}_{0000} = K, \quad
  M^{(4)}_{000(11)} = q^2 K_{\rm I},
\nonumber\\
&
  M^{(4)}_{0011} = M^{(4)}_{0022}
  = -\frac{q^2}{3} K_{\rm I}, \quad
  M^{(4)}_{011(11)} = -q^4 K_{\rm III},
\nonumber\\
&
  M^{(4)}_{022(11)} = q^4 \left(K_{\rm II} - K_{\rm III}\right), \quad
  M^{(4)}_{1111} = M^{(4)}_{2222} = 3q^4 K_{\rm III},
\nonumber\\
&
  M^{(4)}_{1122} =  q^4 K_{\rm III}.
\label{eq:10}
\end{align} 
For other combinations of the indices which cannot be obtained by
permutations in the above equations, $M^{(4)}_{\mu_1\mu_2\mu_3\mu_4} =
0$. Kurtosis parameter and its derivatives are defined by
\begin{align}
&
  K = \frac{\langle f^4 \rangle_{\rm c}}{\sigma^6}, \quad
  K_{\rm I} =
  \frac{\langle f^3 \nabla^2 f \rangle_{\rm c}}
    {\sigma^4\sigma_1^2},
\nonumber\\
&
  K_{\rm II} =
  \frac{2\langle f|\nabla f|^2 \nabla^2 f\rangle_{\rm c}
    + \langle |\nabla f|^4\rangle_{\rm c}}
    {\sigma^2\sigma_1^4}, \quad
  K_{\rm III} =
  \frac{\langle |\nabla f|^4 \rangle_{\rm c}}
    {2\sigma^2\sigma_1^4}.
\label{eq:11}
\end{align}
Substituting Eq.~(\ref{eq:08})--(\ref{eq:10}) into the third term in
the RHS of Eq.~(\ref{eq:06}), we obtain the second-order corrections
to the MFs:
\begin{align}
& v_0^{(2)}(\nu) =
  \frac{S^2}{72} H_5(\nu) + \frac{K}{24} H_3(\nu),
\label{eq:12a}\\
& v_1^{(2)}(\nu) =
  \frac{S^2}{72} H_6(\nu) + \frac{K-S S_{\rm I}}{24} H_4(\nu)
\nonumber\\
& \qquad \qquad \qquad
  - \frac{1}{12}\left(K_{\rm I} + \frac38 S_{\rm I}^2\right) H_2(\nu)
  - \frac{K_{\rm III}}{8},
\label{eq:12b}\\
& v_2^{(2)}(\nu) =
  \frac{S^2}{72} H_7(\nu) + \frac{K-2S S_{\rm I}}{24} H_5(\nu)
\nonumber\\
& \qquad \quad
  - \frac{1}{6}\left(K_{\rm I} + \frac12 S S_{\rm II}\right) H_3(\nu)
  - \frac12 \left(K_{\rm II} + \frac12 S_{\rm I} S_{\rm II}\right)H_1(\nu).
\label{eq:12c}
\end{align}
These equations are our primary results of this paper. In
Eqs.~(\ref{eq:02}), (\ref{eq:03}), (\ref{eq:04}),
(\ref{eq:12a})--(\ref{eq:12c}), the MFs of non-Gaussian fields are
analytically given by the skewness, kurtosis, and their derivatives in
second-order approximations. The above formulas are applicable to any
2D random field which has hierarchical orderings of the higher-order
cumulants.

\section{\label{sec:SK}
Relation to the Bispectrum and Trispectrum
}

Once the skewness, kurtosis and their derivatives are calculated from
a cosmological model, the prediction of MFs follows from our formulas.
The skewness and its derivatives are directly given by the bispectrum,
and the kurtosis and its derivatives are directly given by the
trispectrum.

For a field on a 2D flat space, such as the flat-sky approximation of
the CMB temperature fluctuations $\Delta T(\bm{\theta})/T$, one can
adopt the Fourier transform:
\begin{equation}
  \frac{\Delta T}{T}(\bm{\theta}) = 
  \int\frac{d^2l}{(2\pi)^2} a(\bm{l}) e^{i\bm{l}\cdot\bm{\theta}}.
\label{eq:100-1}
\end{equation}
The Fourier coefficients of the smoothed field $f$ is given by
$\tilde{f}(\bm{l}) = a(\bm{l}) W(l)$, where $W(l)$ is the window
function of the smoothing kernel, where we assume spherically
symmetric smoothing window. For a Gaussian window with smoothing angle
$\theta_{\rm s}$, $W(l) = e^{-l^2\theta_{\rm s}^2/2}$. The power
spectrum $C$, the bispectrum $B$, and the trispectrum $T$ of the 2D
field $\Delta T/T$ are defined by
\begin{align}
&  \langle a(\bm{l}_1) a(\bm{l}_2) \rangle_{\rm c} =
  (2\pi)^2 \delta^2(\bm{l}_1 + \bm{l}_2) C(l_1),
\label{eq:100-2a}\\
&  \langle a(\bm{l}_1) a(\bm{l}_2) a(\bm{l}_3) \rangle_{\rm c}
  = (2\pi)^2 \delta^2(\bm{l}_1 + \bm{l}_2 + \bm{l}_3) B(l_1,l_2,l_3),
\label{eq:100-2b}\\
&  \langle
  a(\bm{l}_1) a(\bm{l}_2) a(\bm{l}_3) a(\bm{l}_4)
  \rangle_{\rm c}
  = (2\pi)^2 \delta^2(\bm{l}_1 + \bm{l}_2 + \bm{l}_3 + \bm{l}_4)
\nonumber\\
& \hspace{10pc} \times
  T(l_1,l_2,l_3,l_4;l_{12},l_{23}),
\label{eq:100-2c}
\end{align}
where $l_i = |\bm{l}_i|$, $l_{ij} = |\bm{l}_i + \bm{l}_j|$. The
translational symmetry is guaranteed by delta functions, and the
rotational symmetry is taken into account in above definitions of
arguments. By straightforward calculations using symmetric
permutations in integrals, one can show that
\begin{align}
  \sigma_{j}^2
  &=
  \int \frac{l_1\,dl_1}{2\pi}\,
  l_1^{2j} C(l_1)W^2(l_1),
\label{eq:101}\\
   S_{A}
   &= \frac{1}{\sigma^4}
   \int \frac{l_1\,dl_1}{2\pi}\frac{l_2\,dl_2}{2\pi} \frac{d\theta_{12}}{2\pi}
    \tilde{S}_{A}(l_1,l_2)
\nonumber\\
&  \hspace{5pc} \times
  B(l_1,l_2,l_{12}) W(l_1) W(l_2) W(l_{12}),
\label{eq:102}\\
  K_A
  &= \frac{1}{\sigma^6}
  \int
  \frac{l_1\,dl_1}{2\pi}\frac{l_2\,dl_2}{2\pi}\frac{l_3\,dl_3}{2\pi}
  \frac{d\theta_{12}}{2\pi}\frac{d\theta_{23}}{2\pi}
  \tilde{K}_A(l_1,l_3,l_{12})
\nonumber\\
& \quad \times
  T(l_1,l_2,l_3,l_4;l_{12},l_{23}) W(l_1) W(l_2) W(l_3) W(l_4),
\label{eq:103}
\end{align}
where
\begin{align}
  l_{12} &= \left(l_1^2 + l_2^2 + 2l_1l_2\cos\theta_{12}\right)^{1/2},
\label{eq:103-1a}\\
  l_{23} &= \left(l_2^2 + l_3^2 + 2l_2l_3\cos\theta_{23}\right)^{1/2},
\label{eq:103-1b}\\
  l_4 &= \left[l_{12}^2 + l_{23}^2 - l_2^2
      + 2l_1l_3\cos(\theta_{12} + \theta_{23}) \right]^{1/2},
\label{eq:103-1c}
\end{align}
and $\sigma_{0} = \sigma$, $(S_A) = (S, S_{\rm I}, S_{\rm
  II})$, $(\tilde{S}_A) = (\tilde{S},\tilde{S}_{\rm I}, \tilde{S}_{\rm II})$,
$(K_A) = (K, K_{\rm I}, K_{\rm II}, K_{\rm III})$, $(\tilde{K}_A) =
(\tilde{K}, \tilde{K}_{\rm I}, \tilde{K}_{\rm II}, \tilde{K}_{\rm III})$,
\begin{align}
&
  \tilde{S} = 1, \quad
  \tilde{S}_{\rm I} = -\frac{l_1^2}{2q^2}, \quad
  \tilde{S}_{\rm II} = 
  \frac{ l_1^2\left(l_{1}^2 - 2l_2^2 \right)}{4q^4},
\label{eq:104a}\\
&
  \tilde{K} = 1, \quad
  \tilde{K}_{\rm I} = -\frac{l_1^2}{2q^2}, \quad
  \tilde{K}_{\rm II} = \frac{l_{12}^4 - 4l_1^2l_3^2}{16q^4},
\label{eq:104b}\\
&
  \tilde{K}_{\rm III} = \frac{l_{12}^4 + 4l_1^2(l_3^2 - l_{12}^2)}{32q^4}.
\label{eq:104c}
\end{align}
The integration ranges are $2 \leq l_i \leq \infty$ and $0 \leq
\theta_{ij} < 2\pi$ when the monopole and dipole components are
subtracted from the map [see Eqs.~(\ref{eq:114a})-(\ref{eq:114c}) and
explanations below].

For a field on a 2D sphere, such as the all-sky CMB map $\Delta
T(\theta,\phi)/T$, one can adopt the expansion by spherical harmonics:
\begin{equation}
  \frac{\Delta T}{T} (\theta,\phi) = \sum_{lm} a_{lm} Y_l^m(\theta,\phi).
\label{eq:105}
\end{equation}
The harmonic coefficients of the smoothed field $f$ is given by
$f_{lm} = a_{lm}W_l$, where $W_l$ is the window function. For a
Gaussian window with smoothing angle $\theta_{\rm s}$, $W_l =
e^{-l(l+1)\theta_{\rm s}^2/2}$. When monopole and dipole components
are subtracted as usually in the case of CMB map, $W_0 = W_1 = 0$. The
power spectrum, bispectrum and trispectrum are defined by
\cite{ver00,kom01,hu01}
\begin{align}
&  \left\langle a_{l_1m_1} a_{l_2m_2} \right\rangle_{\rm c}
  = (-1)^{m_1}\delta_{l_1l_2}\delta_{m_1,-m_2} C_{l_1},
\label{eq:106a}\\
&  \left\langle a_{l_1m_1} a_{l_2m_2} a_{l_3m_3} \right\rangle_{\rm c}
   =\threej{l_1}{l_2}{l_3}{m_1}{m_2}{m_3}B_{l_1l_2l_3},
\label{eq:106b}\\
&  \left\langle a_{l_1m_1} a_{l_2m_2} a_{l_3m_3} a_{l_4m_4}
   \right\rangle_{\rm c}
\nonumber\\
& \qquad
   = \sum_{L,M} (-1)^M \threej{l_1}{l_2}{L}{m_1}{m_2}{-M}
    \threej{l_3}{l_4}{L}{m_3}{m_4}{M} T^{l_1l_2}_{l_3l_4}(L),
\label{eq:106c}
\end{align}
where $\sthreej{l_1}{l_2}{l_3}{m_1}{m_2}{m_3}$ is the Wigner
3$j$-symbol, and the rotational symmetry is taken into account in the
above definitions. The symmetries of the trispectrum
$T^{l_1l_2}_{l_3l_4}(L)$ are enforced by the reduced trispectrum
${\cal T}^{l_1l_2}_{l_3l_4}(L)$ which is an arbitrary function of its
arguments except that it must be symmetric against exchange of its
upper and lower indices ${\cal T}^{l_1l_2}_{l_3l_4}(L) = {\cal
  T}^{l_3l_4}_{l_1l_2}(L)$ \cite{hu01}. The construction is
\begin{multline}
  T^{l_1l_2}_{l_3l_4}(L) = P^{l_1l_2}_{l_3l_4}(L)
  + (2L+1)\sum_{L'}
  \left[
      (-1)^{l_2+l_3}
      \sixj{l_1}{l_2}{L}{l_4}{l_3}{L'} P^{l_1l_3}_{l_2l_4}(L')
  \right.
\\
  \left.
      + (-1)^{L+L'}
      \sixj{l_1}{l_2}{L}{l_3}{l_4}{L'} P^{l_1l_4}_{l_3l_2}(L')
  \right],
\label{eq:107}
\end{multline}
where $\ssixj{l_1}{l_2}{l_3}{l_4}{l_5}{l_6}$ is the Wigner
6$j$-symbol, and
\begin{multline}
  P^{l_1l_2}_{l_3l_4}(L) = 
  {\cal T}^{l_1l_2}_{l_3l_4}(L) +
  (-1)^{l_1+l_2+L} {\cal T}^{l_2l_1}_{l_3l_4}(L)
\\ 
  + (-1)^{l_3+l_4+L} {\cal T}^{l_1l_2}_{l_4l_3}(L)
  + (-1)^{l_1+l_2+l_3+l_4} {\cal T}^{l_2l_1}_{l_4l_3}(L).
\label{eq:108}
\end{multline}

The following properties of spherical harmonics \cite{mes76} are
useful for our purpose:
\begin{align}
&  \nabla^2Y_l^m(\Omega) = -l(l+1)Y_l^m(\Omega),
\label{eq:109a}\\
&  \int d\Omega Y_{l_1}^{m_1}(\Omega)Y_{l_2}^{m_2}(\Omega) = 
  (-1)^{m_1}\delta_{l_1l_2}\delta_{m_1,-m_2},
\label{eq:109b}\\
&  Y_{l_1}^{m_1}(\Omega)Y_{l_2}^{m_2}(\Omega) = 
  \sum_{l_3,m_3}(-1)^{m_3} I_{l_1l_2l_3}
  \threej{l_1}{l_2}{l_3}{m_1}{m_2}{m_3}
  Y_{l_3}^{m_3}(\Omega),
\label{eq:109c}
\end{align}
where
\begin{equation}
  I_{l_1l_2l_3} \equiv
  \sqrt{\frac{(2l_1+1)(2l_2+1)(2l_3+1)}{4\pi}}
  \threej{l_1}{l_2}{l_3}{0}{0}{0}.
\label{eq:110}
\end{equation}
This quantity $I_{l_1l_2l_3}$ is completely symmetric with respect to
permutations of its arguments, non-zero only when $l_1,l_2,l_3$
satisfy the triangular inequality and $l_1+l_2+l_3$ is an even number
\cite{mes76}. With repeated use of
Eqs.~(\ref{eq:109a})--(\ref{eq:109c}), one can calculate the required
parameters. Introducing a notation,
\begin{equation}
  \{l\} \equiv l(l+1),
\label{eq:111-1}
\end{equation}
the results are
\begin{align}
 \sigma_j^2
 &= \frac{1}{4\pi} \sum_l (2l+1) \{l\}^j C_l W_l^2,
\label{eq:111a}\\
  S_{A} &=
  \frac{1}{4\pi \sigma^4} \sum_{l_1,l_2,l_3}
  I_{l_1l_2l_3}
  \tilde{S}_{A\, l_1l_2l_3} B_{l_1l_2l_3} W_{l_1} W_{l_2} W_{l_3},
\label{eq:111b}\\
  K_A &=
  \frac{1}{4\pi \sigma^6} \sum_{l_1,l_2,l_3,l_4,L}
  \frac{I_{l_1l_2L}I_{l_3l_4L}}{2L+1}
  \tilde{K}_{A\,l_3l_4}^{\ \ l_1l_2}(L)
  T^{l_1l_2}_{l_3l_4}(L) W_{l_1} W_{l_2} W_{l_3} W_{l_4},
\label{eq:111c}
\end{align}
where
\begin{align}
&
  \tilde{S}_{l_1l_2l_3} = 1, \quad
  \tilde{S}_{{\rm I}\,l_1l_2l_3} = -\frac{\{l_1\}+\{l_2\}+\{l_3\}}{6q^2},
\label{eq:112a}\\
&
  \tilde{S}_{{\rm II}\,l_1l_2l_3}
  = \frac{\{l_1\}^2 + \{l_2\}^2 + \{l_3\}^2
    - 2 \{l_1\}\{l_2\} - 2 \{l_2\}\{l_3\} - 2 \{l_3\}\{l_1\}}{12q^4},
\label{eq:112b}\\
&
  \tilde{K}^{l_1l_2}_{l_3l_4}(L) = 1, \quad
  \tilde{K}_{{\rm I}\,l_3l_4}^{\ l_1l_2}(L) =
  -\frac{\{l_1\} + \{l_2\} + \{l_3\} + \{l_4\}}{8q^2},
\label{eq:113a}\\
&
  \tilde{K}_{{\rm II}\,l_3l_4}^{\ \ l_1l_2}(L) =
  \frac{\{L\}^2 - \left(\{l_1\} + \{l_2\}\right)
    \left(\{l_3\} + \{l_4\}\right)}{16q^4},
\label{eq:113b}\\
&
  \tilde{K}_{{\rm III}\,l_3l_4}^{\ \ \ l_1l_2}(L) =
  \frac{\left(\{l_1\} + \{l_2\} - \{L\}\right)
    \left(\{l_3\} + \{l_4\} - \{L\}\right)}{32q^4}.
\label{eq:113c}
\end{align}
The above forms of skewness parameters [Eq.~(\ref{eq:112a}),
(\ref{eq:112b})] were already appeared in \cite{hik06}.

Resemblances of the above results to those of the flat space are
obvious if the integrands of Eqs.~(\ref{eq:104a})--(\ref{eq:104c}) are
symmetrized with respect to $l_1, l_2, l_3$, and $l_4$ [conversely,
one can desymmetrize the Eqs.~(\ref{eq:112a})--(\ref{eq:113c}) to have
the similar form with Eqs.~(\ref{eq:104a})--(\ref{eq:104c})]. Noting
the all-sky and flat-sky correspondence \cite{whi99,hu00,hu01}, it is
a straightforward exercise to show that the above all-sky equations
reduce to those of flat-sky in the large-$l$ limit. Following
\cite{whi99,hu00,hu01}, but applying an improved approximation
\begin{equation}
  Y_l^m(\theta,\phi) \approx
  (-1)^m \sqrt{\frac{2l+1}{4\pi}}\ 
  J_m\left[\left(l+\frac12\right)\theta\right] e^{im\phi},
\label{eq:114-1}
\end{equation}
for $\theta \ll 1$, $l \gg 1$, the correspondences between all-sky and
flat-sky spectra are derived as
\begin{align}
&  C_{l} \approx C\left(l+\frac12\right),
\label{eq:114a}\\
&  B_{l_1l_2l_3} \approx I_{l_1l_2l_3}
   B\left(l_1+\frac12,l_2+\frac12,l_3+\frac12\right),
\label{eq:114b}\\
&  P^{l_1l_2}_{l_3l_4}(L) \approx I_{l_1l_2L} I_{l_3l_4L}
\nonumber\\
&\hspace{3pc} \times
   P\left(l_1+\frac12,l_2+\frac12;l_3+\frac12,l_4+\frac12;L+\frac12\right), 
\label{eq:114c}
\end{align}
and
\begin{multline}
  T(l_1,l_2,l_3,l_4;l_{12},l_{23}) = P(l_1,l_2;l_3,l_4;l_{12})
\\
  + P(l_1,l_3;l_2,l_4;l_{13}) + P(l_1,l_4;l_2,l_3;l_{23}),
\label{eq:114d}
\end{multline}
where $l_{13} = [l_1^2 + l_2^2 + l_3^2 + l_4^2 - l_{12}^2 -
l_{23}^2]^{1/2}$.

Since the all-sky multipole $\ell$ and the flat-sky wavenumber
$|\bm{l}|$ are related by $|\bm{l}| = \ell + 1/2$, contributions of
the multipole $\ell$ to the all-sky summation is approximately
represented by the flat-sky integration over the range $\ell-1/2 \leq
|\bm{l}| - 1/2 < \ell + 1/2$, i.e., $\ell \leq |\bm{l}| < \ell + 1$.
Thus, all-sky summations over $\ell = 2,3,\ldots$ correspond to the
flat-sky integrations with the limit $|\bm{l}| \geq 2$, as noted above.
We confirm that the flat-sky approximations of
Eqs.~(\ref{eq:101})--(\ref{eq:103}) with the above correspondences
numerically reproduce the values calculated from all-sky formula of
Eqs.~(\ref{eq:111a})--(\ref{eq:111c}) within several percent for
$\theta_{\rm s} < 100'$.

For numerical evaluations of the kurtosis and its derivatives by the
summation of Eq.~(\ref{eq:111c}), the number of terms to add is of
order ${\cal O}(l_{\rm max}^{\,6})$, where $l_{\rm max}$ is the
maximum multipole required for a given smoothing scale, e.g., $l_{\rm
  max} \sim \mbox{several}\times \theta_{\rm s}^{-1}$. The
computational cost becomes progressively high for large $l_{\rm max}$,
if the summation is naively performed. Efficient evaluations are
necessary when the smoothing angle $\theta_{\rm s}$ is small. In
Appendix~\ref{app:Summation}, numerical schemes for the efficient
evaluations are summarized.

\section{\label{sec:LNG}
A Simple Example:  the Local Model of Non-Gaussianity
}

The analytic MFs are evaluated once the power spectrum, bispectrum and
trispectrum are given. These spectra depend on models of primordial
density fluctuations. As a simple example, we consider below the local
model of non-Gaussianity, although our formulas are not restricted to
this particular model.

In the local model, the primordial curvature perturbations during the
matter era is assumed to take the form
\cite{sal90,ver00,kom01,oka02,kog06}
\begin{equation}
  \Phi(\bm{x}) = \phi(\bm{x})
  + f_{\rm NL} \left(\phi^2(\bm{x}) - \langle\phi^2\rangle\right)
  + g_{\rm NL} \phi^3(\bm{x}),
\label{eq:201}
\end{equation}
in configuration space, where $\phi$ is an auxiliary random Gaussian
field. The comoving curvature perturbation $\zeta$ is given by $\zeta
= 3\Phi/5$. The CMB fluctuations generated by the curvature
perturbations have the harmonic coefficients
\begin{equation}
  a_{lm} = 4\pi (-i)^l \int\frac{d^3k}{(2\pi)^3}
  \tilde{\Phi}(\bm{k}) g^{\rm (T)}_l(k) Y_l^{m*}(\hat{\bm{k}}),
\label{eq:202}
\end{equation}
where $\tilde{\Phi}(\bm{k})$ is the Fourier transform of the
primordial curvature perturbation $\Phi(\bm{x})$, and $g^{\rm
  (T)}_l(k)$ is the radiation transfer function.

The bispectrum and trispectrum of CMB in the local model of
Eq.~(\ref{eq:201}) are derived in literature \cite{kom01,oka02}:
\begin{align}
&
  B_{l_1l_2l_3} = 
  2 f_{\rm NL} I_{l_1l_2l_3}
  \left[
      \int r^2 dr \alpha_{l_1}(r)\beta_{l_2}(r)\beta_{l_3}(r)
      + {\rm cyc.}
  \right],
\label{eq:203a}\\
&
  {\cal T}^{l_1l_2}_{l_3l_4}(L) = I_{l_1l_2L} I_{l_3l_4L}
\nonumber\\
& \quad \times
\left\{
  4 f_{\rm NL}^2
  \int r_1^2dr_1 r_2^2dr_2 F_L(r_1,r_2)
  \alpha_{l_1}(r_1) \beta_{l_2}(r_1)
  \alpha_{l_3}(r_2) \beta_{l_4}(r_2)
\right.
\nonumber\\
& \qquad
\left. +\,
  g_{\rm NL}
   \int r^2 dr \beta_{l_2}(r) \beta_{l_4}(r)
    \left[
        \alpha_{l_1}(r)\beta_{l_3}(r) + \beta_{l_1}(r) \alpha_{l_3}(r)
    \right]
\right\}
\label{eq:203b}
\end{align}
where
\begin{align}
  F_L(r_1,r_2) &\equiv
  4\pi \int \frac{k^2 dk}{2\pi^2} P_\phi(k) j_L(kr_1)  j_L(kr_2),
\label{eq:204a}\\
  \alpha_l(r) &\equiv
  4\pi \int \frac{k^2 dk}{2\pi^2} g^{\rm (T)}_l(k) j_l(kr),
\label{eq:204b}\\
  \beta_l(r) &\equiv
  4\pi \int \frac{k^2 dk}{2\pi^2} P_\phi(k) g^{\rm (T)}_l(k) j_l(kr),
\label{eq:204c}
\end{align}
and $P_\phi(k) \propto k^{n_s-4}$ is the primordial power spectrum of
$\phi$. In some extended models, the factor $4f_{\rm NL}^2$ in the
trispectrum is replaced by $4f_{\rm NL}^2 \rightarrow 25\tau_{\rm
  NL}/9$, and $\tau_{\rm NL}$ is considered as an independent
parameter \cite{byr06}.

The variance parameters $\sigma$ and $\sigma_1$ are also affected by
non-Gaussianity. Therefore we need to evaluate the corrections to the
power spectrum. The non-Gaussian corrections to the parameters
$\sigma$, $\sigma_1$ mostly affect the normalization of MFs. In
practice, the normalization suffers contamination from observational
effects such as pixelization and/or boundary effects, and therefore is
not usually used for extracting cosmological information.
Nevertheless, we include the effect for theoretical consistency here.
By substituting Eq.~(\ref{eq:202}) into Eq.~(\ref{eq:106a}) for a
local model of Eq.~(\ref{eq:201}), performing angular integrations, we
obtain
\begin{multline}
  C_l = 
  \left[ 1 + 6 g_{\rm NL} \int r^2 dr\, \xi_\phi(r)
  \right]
  \tilde{C}_l
\\
  + 2 f_{\rm NL}^2
  \int r^2 dr\, \gamma_l(r) \,\left[\xi_\phi(r)\right]^2,
\label{eq:205}
\end{multline}
where
\begin{align}
  \tilde{C}_l &= 4\pi \int \frac{k^2 dk}{2\pi^2}
  P_\phi(k) \left[g_l^{\rm (T)}(k)\right]^2,
\label{eq:206a}\\
  \xi_\phi(r) &=  \int \frac{k^2 dk}{2\pi^2}
  P_\phi(k) j_0(kr),
\label{eq:206b}\\
  \gamma_l(r) &=  (4\pi)^2 \int \frac{k^2 dk}{2\pi^2}
 \left[g_l^{\rm (T)}(k)\right]^2 j_0(kr).
\label{eq:206c}
\end{align}
The quantities $\tilde{C}_l$, $\xi_\phi(r)$ are the angular power
spectrum from the Gaussian component, and the spatial correlation
function of $\phi$, respectively. Substituting Eqs.~(\ref{eq:203a}),
(\ref{eq:203b}), (\ref{eq:205}) into
Eqs.~(\ref{eq:111a})--(\ref{eq:111c}) with Eqs.~(\ref{eq:107}),
(\ref{eq:108}), all the parameters necessary to calculate the analytic
MFs are evaluated for a model of local non-Gaussianity with a full
transfer function $g^{\rm (T)}_l(k)$.

For the Sachs-Wolfe effect \cite{sac67}, which is valid only
for small multipole moments ($l \ll 100$), the radiation transfer
function is simply given by $g^{\rm (T)}_l(k) = j_l(kr_*)/3$, where
$r_*$ is the comoving distance to the last scattering surface. In this
limit, the reduced bispectrum and trispectrum have the forms
\cite{gan94,kom01,oka02}
\begin{align}
&  B_{l_1l_2l_3}
  = -6 f_{\rm NL}
  \left(
      C^{\rm SW}_{l_1} C^{\rm SW}_{l_2} +
      C^{\rm SW}_{l_2} C^{\rm SW}_{l_3} +
      C^{\rm SW}_{l_3} C^{\rm SW}_{l_1}
   \right) I_{l_1l_2l_3},
\label{eq:207a}\\
&  {\cal T}^{l_1l_2}_{l_3l_4}(L)
  = 9 C^{\rm SW}_{l_2} C^{\rm SW}_{l_4}
\nonumber\\
&\hspace{3.5pc} \times
   \left[
       4 f_{\rm NL}^2 C^{\rm SW}_L
       + g_{\rm NL}\left(C^{\rm SW}_{l_1} + C^{\rm SW}_{l_3}\right)
   \right] I_{l_1l_2L} I_{l_3l_4L},
\label{eq:207b}
\end{align}
where
\begin{equation}
  C^{\rm SW}_l = \frac{2}{9\pi} \int k^2 dk\,
  P_\phi(k) \left[j_l(kr_*)\right]^2,
\label{eq:208}
\end{equation}
is the power spectrum of the Sachs-Wolfe effect from the Gaussian
component. In the same limit, non-Gaussian corrections to the power
spectrum are also obtained after some calculation:
\begin{multline}
  C_l =
  \left[
    1 +  \frac{27g_{\rm NL}}{2\pi}
    \sum_{L}(2L+1)C^{\rm SW}_L
  \right] C^{\rm SW}_l
\\
  + \frac{18 f_{\rm NL}^2}{2l+1}
    \sum_{l_1,l_2} I_{l_1l_2l}^2 C^{\rm SW}_{l_1} C^{\rm SW}_{l_2}.
\label{eq:209}
\end{multline}

\section{\label{sec:Numerical}
Comparisons with Numerical Minkowski Functionals
}

To illustrate how well our formulas work, we compare the analytic MFs
with numerical MFs from realizations of the non-Gaussian CMB map.
Since the purpose of this comparison is to show the performance of
analytic formulas, we simply consider the non-Gaussian map in the
Sachs-Wolfe limit. Implementations with a full radiation transfer
function will be used in future applications to real data. In the
Sachs-Wolfe limit, the temperature fluctuations are given by $\Delta
T_{\rm SW}(\theta,\phi)/T = - \Phi(r_*,\theta,\phi)/3$, and thus
generated by the nonlinear mapping of Gaussian fluctuations $\Delta
T_{\rm G}/T$ at each position in the sky:
\begin{multline}
  \frac{\Delta T_{\rm SW}}{T} =
  \frac{\Delta T_{\rm G}}{T}
  - 3 f_{\rm NL} \left(\frac{\Delta T_{\rm G}}{T}\right)^2
  + 9 g_{\rm NL} \left(\frac{\Delta T_{\rm G}}{T}\right)^3
\\
  - \mbox{(monopole + dipole)},
\label{eq:210}
\end{multline}
where monopole and dipole components are subtracted from the resulting
map. We use the HEALPix package \cite{gor05} to generate the
non-Gaussian map, apply the smoothing window function, and calculate
the first- and second-derivatives of the temperature field. The MFs
are evaluated from the field derivatives using the numerical algorithm
given in \cite{sch98,hik06}. We assume a simple Sachs-Wolfe power
spectrum of $l(l+1)C^{\rm SW}_l/2\pi = 10^{-10}$ for $l \leq 128$ and
$C^{\rm SW}_l=0$, otherwise. Without this cut-off, the resulting power
spectrum would logarithmically diverge in Eq.~(\ref{eq:209}) (the
actual power spectrum has natural damping in high-$l$ regime). We
adopt the Gaussian filter $W_l = e^{-l(l+1)\theta_{\rm s}^2/2}$ with
the smoothing radius of $\theta_{\rm s} = 100'$, and non-Gaussianity
parameters $f_{\rm NL} = 10^2$ and $g_{\rm NL} = 10^6$. We generate
$100,000$ realizations of the non-Gaussian map, numerically calculate
MFs in each realization, and finally average over the realizations.

\begin{figure}
  \includegraphics[width=19.5pc]{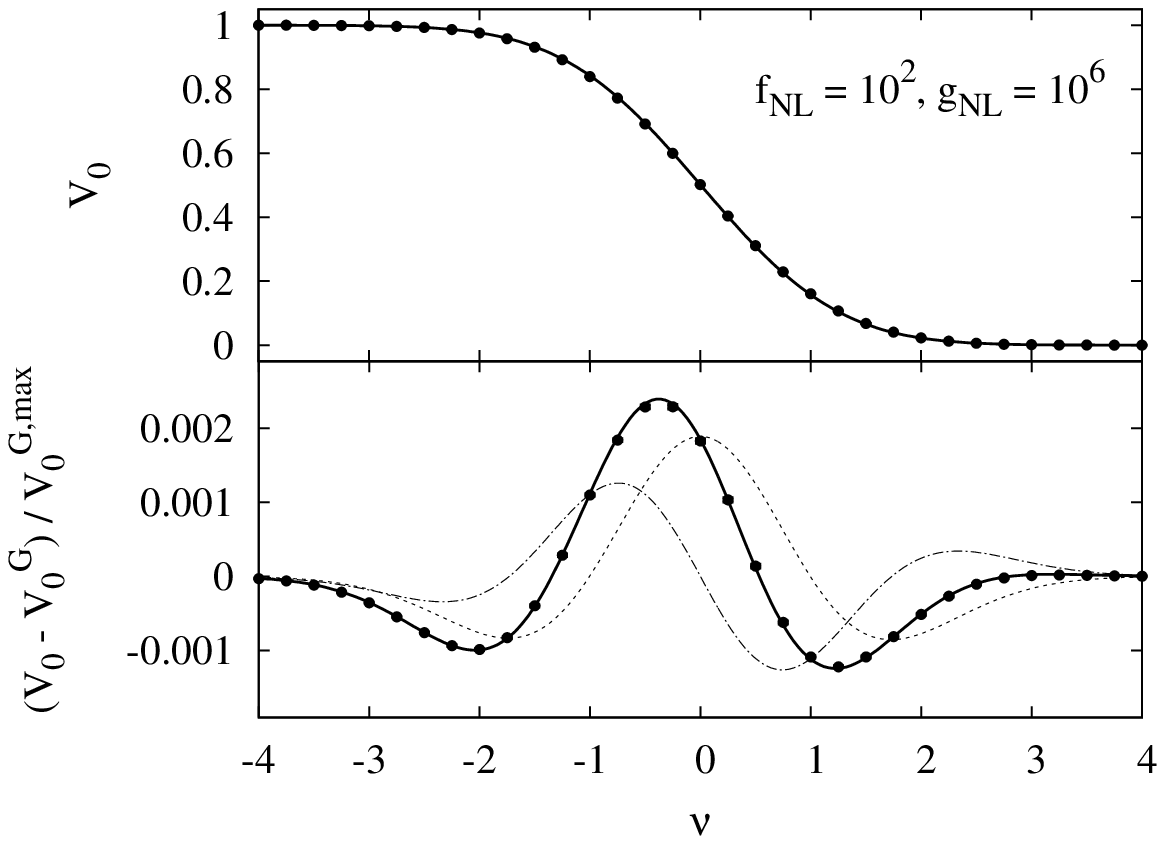}
  \includegraphics[width=19.5pc]{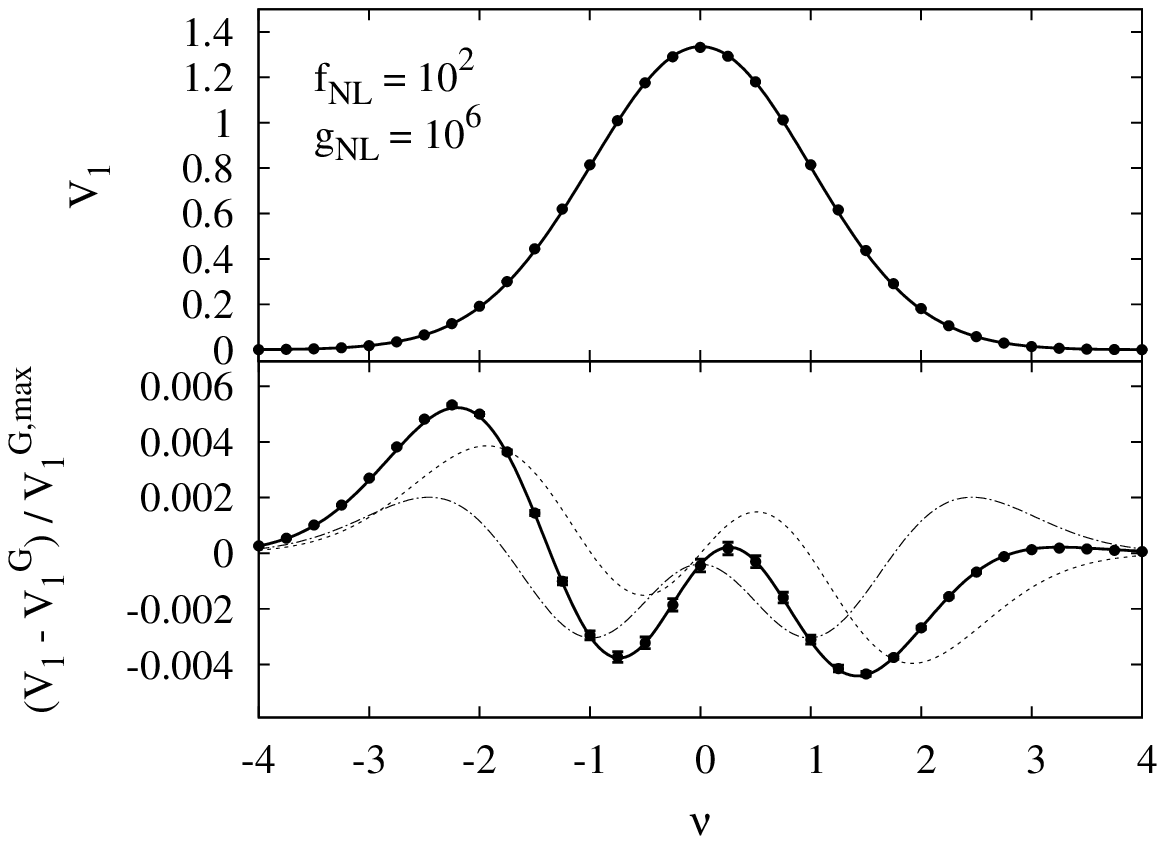}
  \includegraphics[width=19.5pc]{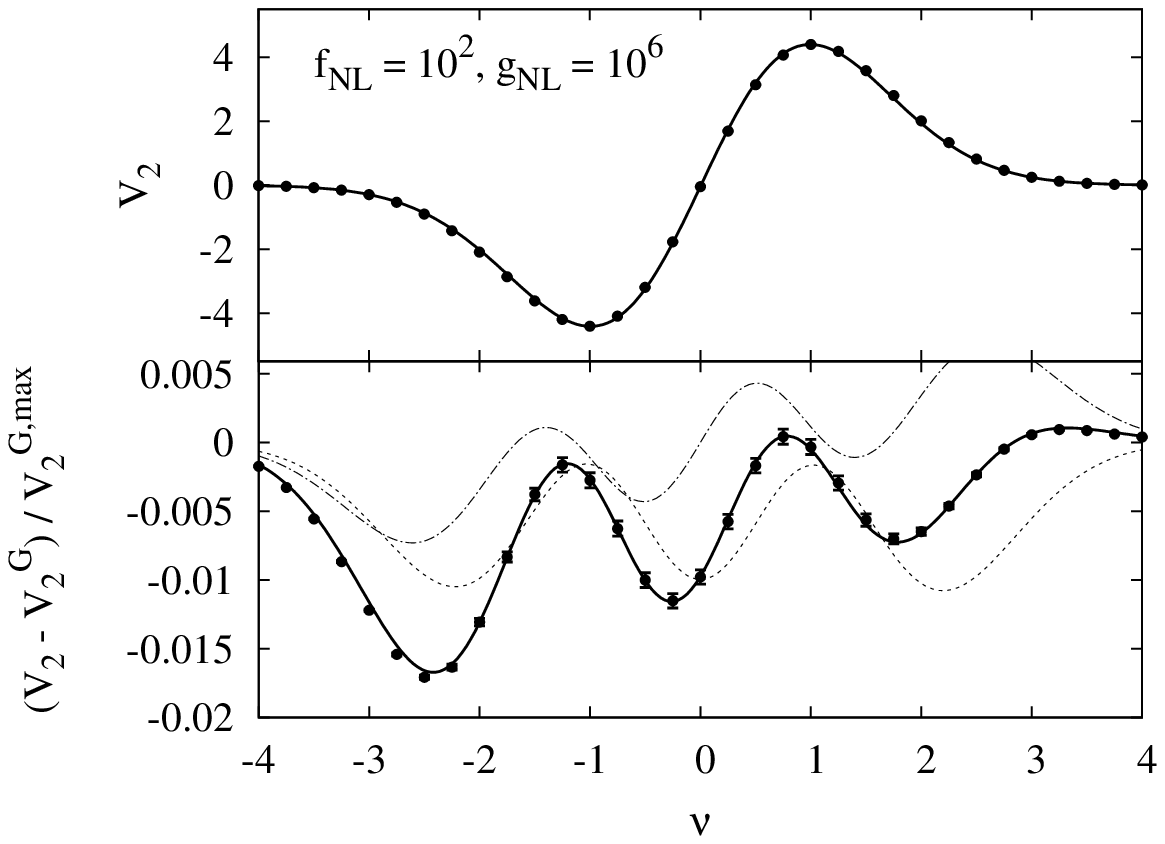}
  \caption{Comparisons between analytic MFs (solid lines) and
    numerical MFs (symbols) in the Sachs-Wolfe limit. We use
    $\theta_{\rm s} = 100'$ and $f_{\rm NL} = 10^2$, $g_{\rm NL} =
    10^6$. Contributions from $f_{\rm NL}$ (dotted lines) and
    $g_{\rm NL}$ (dot-dashed lines) are also shown. Total MFs are
    given by sums of the two contributions in such a weak
    non-Gaussianity regime.
    \label{fig:01}}
\end{figure}
In Fig.~\ref{fig:01}, the analytic and numerical MFs are compared. MFs
as functions of the threshold $\nu$ are plotted in upper panels. As
noted in \cite{hik06}, precise values of overall normalizations are
affected by numerical artifacts such as pixelization, boundary effects
etc.~even in the Gaussian random field at the sub-percent level.
However, deviations from the Gaussian shape of MFs as functions of the
threshold $\nu$ are in very good agreement between numerical and
analytic MFs. In lower panels, differences between non-Gaussian MFs
and Gaussian MFs, divided by the maximum amplitudes, are compared. For
numerical data, MFs from Gaussian realizations are calculated and
subtracted from those of non-Gaussian realizations. We find the
agreements are extremely well. Note that there is no fitting parameter
at all in the comparisons.

Contributions from the parameters $f_{\rm NL}$ and $g_{\rm NL}$ are
shown in thin dotted lines and dot-dashed lines, respectively. Phases
of oscillating patterns are different from one another. This is
because the higher-order non-Gaussianity involves higher-order Hermite
polynomials as given in Eqs.~(\ref{eq:04}) and
(\ref{eq:12a})--(\ref{eq:12c}). Since the Hermite polynomials are
orthogonal functions, the effect of each parameter can be
distinguished with such observations.

\section{Summary
\label{sec:concl}
}

In this paper, the analytic MFs with second-order non-Gaussianity,
including effects from the bispectrum and trispectrum, have been
derived. As long as the higher-order cumulants obey hierarchical
structure, $\langle f^n \rangle_{\rm c} \sim {\cal O}(\sigma^{2n-2})$,
deviations from the Gaussian predictions of MFs are expanded by
$\sigma$. First several cumulants are relevant to the non-Gaussian
corrections. Up to the second-order in $\sigma$, the non-Gaussian
corrections are expressed by the skewness, kurtosis, and their
derivatives, and thus by the bispectrum and trispectrum. These
fundamental results are general and applicable to any 2D random field
which has hierarchical orderings of higher-order cumulants. In
cosmology, applications to the CMB temperature fluctuations are of
great importance. Once the form of bispectrum and trispectrum in CMB
are given by a model of the early universe, the analytic MFs are
evaluated by our formulas. We have illustrated how the analytic MFs of
CMB are calculated, employing the local model of non-Gaussianity, as
an example. We have compared the analytical MFs with numerical
calculations of non-Gaussian MFs, and found very good agreements.
Applications to real data of CMB map are now in progress and will be
reported in future work.

\begin{acknowledgments}
    I wish to thank C.~Hikage for discussion and comments. I
    acknowledge support from the Ministry of Education, Culture,
    Sports, Science, and Technology, Grant-in-Aid for Scientific
    Research (C), 21540263, 2009, and Grant-in-Aid for Scientific
    Research on Priority Areas No. 467 ``Probing the Dark Energy
    through an Extremely Wide and Deep Survey with Subaru Telescope.''
    This work is supported in part by JSPS (Japan Society for
    Promotion of Science) Core-to-Core Program ``International
    Research Network for Dark Energy.''
\end{acknowledgments}

\bigskip
\appendix

\section{\label{app:Summation}
Evaluations of Skewness, Kurtosis and Their Derivatives
}

When the Eq.~(\ref{eq:111c}) is straightforwardly evaluated, the
number of terms to add is of order ${\cal O}(l_{\rm max}^{\,6})$,
where $l_{\rm max}$ is the maximum multipole moment required for a
given smoothing scale, e.g., $l_{\rm max} \sim
\mbox{several}\times\theta_{\rm s}^{-1}$. For parameters $K$ and
$K_{\rm I}$, summations over $L$ for terms with 6$j$-symbols can be
analytically performed, but we still need numerical summations of
6$j$-symbols for parameters $K_{\rm II}$ and $K_{\rm III}$. The
computational cost is progressively high for large $l_{\rm max}$, and
therefore efficient calculations are necessary when the smoothing
angle is small.

First of all, one should take advantages of symmetries and constraints
to save the number of additions. For the skewness parameters, the
number of addition is reduced by a factor of about six, using the
substitution
\begin{equation}
  \sum_{l_1,l_2,l_3} = \sum_{l_1=l_2=l_3} +\, 3\sum_{l_1=l_2<l_3} +\,
  3\sum_{l_1<l_2=l_3} +\, 6\sum_{l_1<l_2<l_3},
\label{eq:a02}
\end{equation}
in Eq.~(\ref{eq:111b}). The summation is taken only when $l_1+l_2+l_3$
is an even number and $l_1,l_2,l_3$ satisfy the triangular inequality,
$l_2 - l_1 \leq l_3 \leq l_1 + l_2$ because of the factor $I_{l_1l_2l_3}$.
For the kurtosis parameters, the number of addition is reduced by a
factor of about eight, using the substitution
\begin{multline}
  \sum_{l_1,l_2,l_3,l_4}
  =
  \sumst{l_1=l_2}{l_3=l_4}{l_1=l_3}
  +\, 2\sumst{l_1=l_2}{l_3=l_4}{l_1<l_3}
  +\, 2\sumst{l_1=l_2}{l_3<l_4}{l_1=l_3}
  +\, 2\sumst{l_1<l_2}{l_3=l_4}{l_1=l_3}\\
  +\, 4\sumst{l_1=l_2}{l_3<l_4}{l_1<l_3}
  +\, 4\sumst{l_1<l_2}{l_3=l_4}{l_1<l_3}
  +\, 4\sumst{l_1<l_2}{l_3<l_4}{l_1=l_3}
  +\, 8\sumst{l_1<l_2}{l_3<l_4}{l_1<l_3},
\label{eq:a04}
\end{multline}
in Eq.~(\ref{eq:111c}). The summation is taken only when
$l_1+l_2+l_3+l_4$ is an even number and $l_4 \leq l_1 + l_2 + l_3$ is
satisfied because of the factor $I_{l_1l_2L}I_{l_3l_4L}$. The index
$L$ in Eq.~(\ref{eq:111c}) runs over every other integers in the range
$[{\rm max}(|l_1-l_2|,|l_3-l_4|),{ \rm min}(l_1+l_2,l_3+l_4)]$ with
the same parity as that of $l_1+l_2$.

Secondly, since the trispectrum is not a strongly oscillating function
in usual cases, one can sparsely sample the multipoles in the sum for
large values of $l_1,\ldots,l_4$. Acoustic oscillations in the CMB are
sufficiently mild in this respect. There is a caveat that the
summation over $L'$ in Eq.~(\ref{eq:107}) should not be sparsely
sampled, since the 6$j$-symbol is a strongly oscillating function with
respect to its arguments. Sparse samplings for large $l_i$'s with
appropriate weights save enormous amounts of time for the calculation.

Instead of calculating the exact summation, one can use the flat-sky
approximation of Eq.~(\ref{eq:103}), and evaluate the five-dimensional
integral by, e.g., the Monte-Carlo algorithm. Once the reduced
trispectrum ${\cal T}^{l_1l_2}_{l_3l_4}(L)$ is given,
Eqs.~(\ref{eq:108}), (\ref{eq:114c}), (\ref{eq:114d}) give the
trispectrum $T(l_1,l_2,l_3,l_4;l_{12},l_{23})$ in the flat-sky
approximation. There is no need for summing 6$j$-symbols in this
procedure. The integration ranges of Eq.~(\ref{eq:103}) are $2 \leq
l_1, l_2, l_3, l_4 < \infty$, $0 \leq \theta_{12}, \theta_{23} \leq
2\pi$ when the monopole and dipole components are subtracted from the
map. This approximation is valid when most contributions to the
kurtosis parameters come from the flat-sky regime ($l \gg 1$). In
realistic spectra, this condition is satisfied because multipoles near
the acoustic peak at $l\sim 200$ dominantly contribute. Strictly
speaking, the flat-sky integration is valid when all $l$'s are large.
Therefore, for a more precise approximation, one can use the all-sky
summation of Eq.~(\ref{eq:111c}) only when, e.g., $l_1 \leq 20$ [note
that $l_1 \leq l_2,l_3,l_4$ in Eq.~(\ref{eq:a04})], and otherwise
adopt the flat-sky integration of Eq.~(\ref{eq:103}) over the range
$l_1,l_2,l_3,l_4 \geq 21$.

Unless sub-percent level accuracies are required, experiences show
that the simplest flat-sky integrals over all ranges of multipole with
appropriate choice of integration limits suffice for evaluations of
skewness and kurtosis parameters in realistic spectra.

\newcommand{\aap}{Astron. Astrophys. }
\newcommand{\apjl}{Astrophys. J. Letters }
\newcommand{\apjs}{Astrophys. J. Suppl. Ser. }
\newcommand{\apss}{Astrophys. Space Sci. }
\newcommand{\jcap}{J. Cosmol. Astropart. Phys. }
\newcommand{\mnras}{Mon. Not. R. Astron. Soc. }
\newcommand{\mpla}{Mod. Phys. Lett. A }
\newcommand{\pasj}{Publ. Astron. Soc. Japan }
\newcommand{\physrep}{Phys. Rep. }
\newcommand{\ptp}{Progr. Theor. Phys. }


\begin{thebibliography}{10}

\bibitem{muk81} V.~F.~Mukhanov and G.~V.~Chibisov, Soviet Journal of
    Experimental and Theoretical Physics Letters {\bf 33}, 532 (1981).
\bibitem{sta82} A.~A.~Starobinsky, \pl B {\bf 117}, 175
    (1982).
\bibitem{haw82} S.~W.~Hawking, \pl B {\bf 115}, 295 (1982).
\bibitem{gut82} A.~H.~Guth and S.-Y.~Pi, \prl {\bf 49}, 1110 (1982).

\bibitem{all87} T.~J.~Allen, B.~Grinstein and M.~B.~Wise, \pl B {\bf
      197}, 66 (1987).
\bibitem{gan94} A.~Gangui, F.~Lucchin, S.~Matarrese, and S.~Mollerach,
    \apj {\bf 430}, 447 (1994).

\bibitem{lyt02} D.~H.~Lyth and D.~Wands, Phys. Lett. B {\bf 524}, 5 (2002).
\bibitem{lin97} A.~Linde and V.~Mukhanov, \prd {\bf 56}, R535 (1997).
\bibitem{lyt03} D.~H.~Lyth, C.~Ungarelli and D.~Wands,  \prd {\bf
      67}, 023503 (2003). 
\bibitem{arm99} C.~Armend{\'a}riz-Pic{\'o}n, T.~Damour and
    V.~Mukhanov, \pl B {\bf 458}, 209 (1999).
\bibitem{gar99} J.~Garriga and V.~F.~Mukhanov, \pl B {\bf 458}, 219
    (1999).
\bibitem{ali04} M.~Alishahiha, E.~Silverstein and D.~Tong, \prd {\bf
      70}, 123505 (2004).

\bibitem{kho01} J.~Khoury, B.~A.~Ovrut,  P.~J.~Steinhardt and
    N.~Turok, \prd {\bf 64}, 123522 (2001).
\bibitem{bar04} N.~Bartolo, E.~Komatsu, S.~Matarrese and A.~Riotto,
    \physrep {\bf 402}, 103 (2004).
\bibitem{kom03} E.~Komatsu, et al., \apjs {\bf 148}, 119 (2003).
\bibitem{spe07} D.~N.~Spergel, et al., \apjs {\bf 170}, 377 (2007).
\bibitem{cre07} P.~Creminelli, L.~Senatore, M.~Zaldarriaga and
M.~Tegmark, \jcap {\bf 3}, 5 (2007).
\bibitem{kom09} E.~Komatsu, et al., \apjs {\bf 180}, 330 (2009).
\bibitem{smi09} K.~M.~Smith, L.~Senatore and M.~Zaldarriaga, \jcap
    {\bf 9}, 6 (2009).
\bibitem{cur09} A.~Curto, E.~Mart{\'{\i}}nez-Gonz{\'a}lez,
    P.~Mukherjee, R.~B.~Barreiro, F.~K.~Hansen, M.~Liguori and
    S.~Matarrese, \mnras {\bf 393}, 615 (2009).
\bibitem{sas06} M.~Sasaki, J.~V{\"a}liviita and D.~Wands, \prd {\bf
      74}, 103003 (2006).
\bibitem{hua09} Q.-G.~Huang, \jcap {\bf 6}, 35 (2009).
\bibitem{byr09} C.~T.~Byrnes and G.~Tasinato, \jcap {\bf 8}, 16 (2009).
\bibitem{leh09} J.-L.~Lehners, S.~Renaux-Petel, \prd {\bf 80}, 063503
    (2009).
\bibitem{mec94} K.~R.~Mecke, T.~Buchert and H.~Wagner, \aap {\bf
      288}, 697 (1994).
\bibitem{sch97} J.~Schmalzing, and T.~Buchert, \apjl {\bf 482}, L1
    (1997).
\bibitem{sch98} J.~Schmalzing, and K.~M.~Gorski, \mnras {\bf 297},
  355 (1998).
\bibitem{nov00} D.~Novikov, J.~Schmalzing V.~F.~Mukhanov, \aap {\bf
      364}, 17 (2000).
\bibitem{wu01} J.~H.~P.~Wu, A.~Balbi, J.~Borrill, P.~G.~Ferreira,
    S.~Hanany, A.~H.~Jaffe, A.~T.~Lee, B.~Rabii, P.~L.~Richards,
    G.~F.~Smoot, R.~Stompor and C.~D.~Winant, \prl {\bf 87}, 251303
    (2001).
\bibitem{pol02} G.~Polenta, et al., \apjl {\bf 572}, L27 (2002).
\bibitem{cur07} A.~Curto, J.~Aumont, J.~F.~Mac\'ias-P\'erez,
    E.~Mart\'inez-Gonz\'alez, R.~B.~Barreiro, D.~Santos,
    F.~X.~D\'esert and M.~Tristram, \aap {\bf 474}, 23 (2007).
\bibitem{det07} G.~de~Troia, et al., \apjl {\bf 670}, L73 (2007).
\bibitem{cur08} A.~Curto, J.~F.~Mac\'ias-P\'erez,
    E.~Mart\'inez-Gonz\'alez, R.~B.~Barreiro, D.~Santos, F.~K.~Hansen,
    M.~Liguori and S.~Matarrese, \aap {\bf 486}, 383 (2008).

\bibitem{mat03} T.~Matsubara, \apj {\bf 584}, 1 (2003)

\bibitem{hik06} C.~Hikage, E.~Komatsu, and T.~Matsubara, \apj {\bf
      653}, 11 (2006).

\bibitem{hik08} C.~Hikage, T.~Matsubara, P.~Coles, M.~Liguori,
    F.~K.~Hansen and S.~Matarrese, \mnras {\bf 389}, 1439 (2008).
\bibitem{nat09} P.~Natoli, et al., arXiv:0905.4301 (2009).
\bibitem{hik09} C.~Hikage, K.~Koyama, T.~Matsubara, T.~Takahashi,
    M.~Yamaguchi, \mnras {\bf 398}, 2188 (2009).

\bibitem{chi09} P.~Chingangbam, and, C.~Park, Journal of Cosmology and
    Astro-Particle Physics {\bf 12}, 19 (2009).
\bibitem{pog09} D.~Pogosyan, C.~Gay, and C.~Pichon, \prd {\bf 80},
    081301(R) (2009). 


\bibitem{tom86} H.~Tomita, \ptp {\bf 76}, 952 (1986).

\bibitem{ver00} L.~Verde, L.~Wang, A.~F.~Heavens, and M.~Kamionkowski,
    \mnras {\bf 313}, 141 (2000).
\bibitem{kom01} E.~Komatsu, and D.~N.~Spergel, \prd {\bf 63}, 063002
    (2001).
\bibitem{hu01} W.~Hu, \prd {\bf 64}, 083005 (2001).

\bibitem{mes76} A.~Messiah, {\it Quantum Mechanics}, Vol.2 (Amsterdam:
    North-Holland, 1976)

\bibitem{hu00} W.~Hu, \prd {\bf 62}, 043007 (2000).

\bibitem{whi99} M.~White, J.~E.~Carlstrom, 
M.~Dragovan, and W.~L.~Holzapfel, \apj {\bf 514}, 12 (1999).

\bibitem{sal90} D.~S.~Salopek, and J.~R.~Bond, \prd {\bf 42}, 3936
    (1990).

\bibitem{oka02} T.~Okamoto, and W.~Hu, \prd {\bf 66}, 063008 (2002).

\bibitem{kog06} N.~Kogo, and E.~Komatsu, \prd {\bf 73}, 083007 (2006).

\bibitem{byr06} C.~T.~Byrnes, M.~Sasaki and D.~Wands, \prd {\bf 74},
    123519 (2006).

\bibitem{sac67} R.~K.~Sachs, and A.~M.~Wolfe, \apj {\bf 147}, 73 (1967).

\bibitem{gor05} K.~M.~G\'orski, E.~Hivon, A.~J.~Banday, B.~D.~Wandelt,
    F.~K.~Hansen, M.~Reinecke, and M~Bartelmann, \apj {\bf 622},759
    (2005).




\end{thebibliography}

\end{document}